\documentclass[reprint,superscriptaddress,preprintnumbers,twocolumn,groupedaddress,showpacs,nofootinbib]{revtex4-1}
\usepackage{graphicx}
\usepackage{ulem}
\usepackage{amsmath, amsthm, amssymb}
\usepackage{amsfonts}
\usepackage{subfig}
\usepackage{xspace}
\usepackage{paralist}
\usepackage{multirow}
\usepackage{paralist}
\usepackage{graphicx}
\usepackage{bm}
\usepackage{times}
\usepackage{slashed}
\usepackage{color}
\usepackage{epsfig}\usepackage{dcolumn}

\usepackage[colorlinks,pdfstartview=FitH]{hyperref}
\hypersetup{linkcolor=blue,citecolor=blue,filecolor=black,urlcolor=blue}

\usepackage{amsmath,amssymb,bm}
\usepackage{slashed}
\usepackage{graphicx}
\usepackage{amsfonts}
\usepackage{color}

\usepackage{amsfonts}
\usepackage{bbm}
\usepackage{latexsym}

\begin{document}

\title{A try for dark energy in quantum field theory: The vacuum energy of neutrino field}

\author{Lian-Bao Jia}  \email{jialb@mail.nankai.edu.cn}
\affiliation{
School of Mathematics and Physics, Southwest University of Science and Technology, Mianyang 621010, China}

\begin{abstract}

The quartic-divergent vacuum energy poses an ultraviolet (UV) challenge (the cosmological constant problem) in probing the nature of dark energy. Here we try to evaluate the contribution of the vacuum energy to dark energy in the UV-free scheme. The result indicates that it is not a problem of a field in the UV region but a question of the contributions of heavy fields being suppressed. Then, we explore an effective description via scale decoupling. The parameter spaces suggest that the vacuum energy of active neutrino fields can naturally meet the observation of dark energy density, and a neutrino with a typical mass $\sim$ 10 meV $(10^{-3}$ eV) is expected. The normal ordering neutrinos are preferred by naturalness, and the neutrino mass window set by dark energy is 6.3 meV $\lesssim m_1 \lesssim$ 16.3 meV, 10.7 meV $\lesssim m_2 \lesssim$ 18.4 meV, 50.5 meV $\lesssim m_3 \lesssim$ 52.7 meV.

\end{abstract}

	\maketitle
	
	\newpage

\section{Introduction}

Physics thrives on crisis \cite{Weinberg:1988cp}, and the ultraviolet (UV) problem is one of the keys in the development of modern physics. The UV catastrophe in classical blackbody radiation led Planck to introduce energy quanta \cite{Planck:1901tja}. In quantum field theory (QFT), the UV problem reappears, that is, loop corrections of a transition process are often UV divergent. The transition amplitude obtained by Feynman rules (the physical input) is not directly equal to the physical result (the physical output). To extract the finite result from a UV-divergent input, the traditional approach is regularization \& renormalization, that is, the UV divergence is firstly expressed by the regulator based on equivalent transformation and then removed by counterterms, with the mathematical structure of this route being $\infty - \infty$. This traditional approach is successful in dealing with logarithmic divergences in the standard model (SM). When we leave the logarithmic region and move on, there are UV problems for power-law divergences, i.e., \\
(a) The hierarchy problem of the Higgs mass. Loop corrections to the Higgs mass are power-law divergences, with the fine-tuning of the 125 GeV Higgs \cite{tHooft:1979rat,Susskind:1978ms,Branchina:2022jqc,Branchina:2022gll}. \\
(b) The non-renormalizable Einstein gravity. When quantizing Einstein gravity, graviton loops exhibit power-law divergences, requiring an infinite number of counterterms \cite{DeWitt:1967yk,tHooft:1974toh,Goroff:1985th}. \\
(c) The cosmological constant problem. The vacuum energy density in QFT is quartically divergent. At a Planck-scale cutoff, its contribution exceeds the critical density for a flat universe by about 120 orders of magnitude \cite{Weinberg:1988cp,SupernovaSearchTeam:1998fmf,SupernovaCosmologyProject:1998vns,Peebles:2002gy,Padmanabhan:2002ji,Copeland:2006wr,Li:2011sd}, with the observed dark energy density $\sim (2.3\,\mathrm{meV})^4$.

How do we deal with the above three UV problems? The traditional approach struggles with power-law divergences, requiring mathematically precarious subtractions via counterterms. Indeed, Dirac \cite{DiracTheEO} viewed renormalization as mathematically provisional, noting that the physical laws chosen by nature possess mathematical beauty. Feynman \cite{Feynman:1986er} stressed the need for mathematical self-consistency and rigor. Dyson \cite{Dyson:1949ha} and Schwinger \cite{Schwinger} envisioned future loop corrections based on finite quantities. Wilsonian effective theory \cite{Wilson:1971bg} decouples scales via a cutoff. Hence, is it not time to explore an intrinsically finite route? Such a route first reproduces the renormalization results for logarithmic divergences, then addresses the UV problems of power-law divergences.

For loops, assuming the physical contributions are insensitive to UV regions, a method of the UV-free scheme \cite{Jia:2023dub} is introduced to derive loop contributions, i.e., it implements the physical locality of scales inherent in effective theories such as the SM as they evolve with energy scale. It takes a route of analytic continuation from the physical input $\mathcal{T}_\mathrm{F}$ to the physical output $\mathcal{T}_\mathrm{P}$ in a mathematical structure $\mathcal{T}_\mathrm{F} \to \mathcal{T}_\mathrm{P}$,
\begin{eqnarray}
 \mathrm{Input} \bigg\{ \!\! \begin{array}{c}
  \textit{Equivalent transformation} \, , \infty - \infty \\
  \textit{Analytic continuation} \, , \,\, \mathcal{T}_\mathrm{F} \to \mathcal{T}_\mathrm{P} ~~~
\end{array} \!\! \bigg\}  \mathrm{Output} \, , \nonumber
\end{eqnarray}
with UV divergences systematically eliminated without counterterms, thus removing the need for UV subtractions. Loop contributions constructed through finite-valued quantities are free of UV divergences, as guaranteed by locality-based analytic continuation.

From a computational perspective, the UV-free scheme can be understood simply. Using finite-valued quantities in the Lagrangian, one introduces an auxiliary parameter $\xi$ into the denominator of a propagator, differentiates the amplitude a sufficient number of times to make the loop integral UV convergent, integrates over the loop momentum, takes the antiderivative with respect to $\xi$, and finally takes the limit $\xi\to 0$. The boundary constant $C$ is fixed by physical conditions. This preserving-locality method, alongside the Feynman rules, directly yields the physical amplitude $\mathcal{T}_\mathrm{P}$ (no UV divergence in calculations, hence no bare infinity introduced for UV subtractions, unlike traditional renormalization).

This method reproduces the renormalization results for logarithmic divergences. For power-law divergences, this method interprets the Higgs mass hierarchy problem within SM \cite{Jia:2023dub}, and an application to Einstein gravity was investigated in Ref. \cite{Jia:2024lyy}. Here we focus on the UV-divergent vacuum energy. In exploring the nature of dark energy, a central and unavoidable question is the vacuum energy density $\rho_\mathrm{vac}$ that corresponds to a cosmological constant ($\Lambda = 8 \pi G \rho_\mathrm{vac}$). For the cosmological constant problem, a viable path must first identify vacuum energy's physical contribution, presently obscured by UV divergences. The introduction of new physics, whether as particles or fields, becomes justifiable only after the clarification reveals an insufficiency. Since the form of the vacuum energy ($\varpropto \! \int \!\! \frac{d^3k}{(2\pi)^3} \sqrt{k^2\!+\!m^2}$) resembles a loop integral, the UV-free scheme will be employed as a uniform treatment.

\section{The vacuum energy in QFT}

Dark energy plays an important role in cosmic evolution, with equation of state $p_\mathrm{vac}=\omega \rho_\mathrm{vac}$ ($\omega =-1$). In QFT, the vacuum energy density $\rho_\mathrm{vac}$ can be obtained by summing the zero-point energy of all fields. For a field $i$ with a mass $m_i$, its contribution is
\begin{eqnarray} \label{vacE}
\rho_0^i = (-1)^{2 j_i} g_i \int \!\! \frac{d^3k}{(2\pi)^3} \frac{1}{2}\sqrt{k^2+m_i^2} \, ,
\end{eqnarray}
where $j_i$ is the spin of the field, and $g_i$ is the number of degrees of freedom. This is a quartic divergence, and not a loop contribution, independent of coupling parameters (no bare parameter to absorb it via $\infty - \infty$, no coupling to run; a hard Planck-scale cutoff is the standard recourse within the traditional approach). The UV-free scheme \cite{Jia:2023dub,Jia:2024lyy}, a physical analytic continuation of the UV-divergent integral, is adopted to evaluate the contribution of Eq. (\ref{vacE}). The Feynman amplitude $\mathcal{T}_\mathrm{F}^i = \rho_0^i$ is the physical input, and the Feynman-like amplitude $\mathcal{T}_\mathrm{F}^i (\xi)$ here can be obtained with $\sqrt{k^2\!+\!m_i^2}$ replaced by $\sqrt{k^2\!+\!m_i^2 \!+\!\xi}$. The physical transition amplitude $\mathcal{T}_\mathrm{P}^i$ is
\begin{eqnarray}
\mathcal{T}_\mathrm{P}^{i} \! & = & \! \Big [  \mathrm{\int} (d\xi)^3 \frac{\partial^3  \mathcal{T}_\mathrm{F}^i (\xi)}{\partial \xi^3} \Big]_{\xi \to 0} \!\! + \! C       \\
&=& \! \Big [  (-1)^{2 j_i} g_i \! \mathrm{\int} \! (d\xi)^3  \!\! \int \!\! \frac{d^3k}{(2\pi)^3} \frac{3}{16}\frac{1}{\big(\sqrt{k^2\!+\!m_i^2\!+\!\xi}\big)^5} \Big]_{\xi \to 0} \!\! + \! C  \nonumber \\
&=&   (-1)^{2 j_i} g_i \frac{1}{64 \pi^2} m_i^4 \log (m_i^2)  +  C\,  \,    .             \nonumber
\end{eqnarray}
The result with $C$ set by a reference energy scale $\mu_\Lambda^{}$ is
\begin{eqnarray} \label{neutrinoE}
\mathcal{T}_\mathrm{P}^{i} =  (-1)^{2 j_i} g_i \frac{1}{64 \pi^2} m_i^4 \log \frac{m_i^2}{\mu_\Lambda^2}  \,   .
\end{eqnarray}
The Hamiltonian zero-point energy is not manifestly Lorentz invariant, while the UV-free scheme yields a finite density that is a Lorentz scalar, making the quartically divergent zero-point energy of free space well-defined in this non-perturbative context.\footnote{For another type UV-divergence caused by the discrete sequence, i.e., the Casimir effect from bounded vacuum field modes, the Riemann zeta function $\zeta (-1)=-\frac{1}{12}$ can be adopted as the analytic continuation.} In the limit $m_i \to 0$, the massless field gives zero, just as the one-loop gluon seagull diagram does. The scale $\mu_\Lambda$ arises in odd-dimensional spaces, but not in even ones. It characterizes the vacuum energy, analogous to $\Lambda_\mathrm{QCD}$ for the strong interaction and $v$ for the electroweak sector.

Here is an interpretation of scales in QFT. In a scattering process, the full physical result (tree and loop levels) is independent of the free parameter $\mu$ used in perturbative expansions; it can be equivalently described by a tree-level process with an effective coupling at the scale $\mu_\mathrm{eff}$. The vacuum energy scale $\mu_\Lambda^{}$ is analogous to this physical scale $\mu_\mathrm{eff}$. Unlike $\mu_{\mathrm{eff}}$, which varies with scattering energy, $\mu_\Lambda^{}$ acts as a common reference scale for all field ground states, and may be fixed or slowly running with cosmic evolution.

\section{The ripple description and neutrino field}

\begin{figure}[htbp!]
\includegraphics[width=0.2\textwidth]{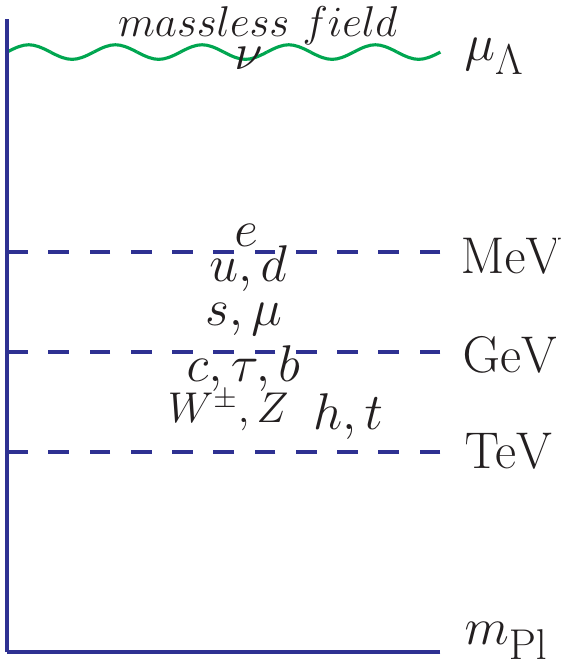}
\caption{The vacuum energy in the ripple description.}
\label{ripple1}
\end{figure}

The QFT vacuum in free space contains zero-point fluctuations from the uncertainty principle. Their short-distance energy density is given by Eq. (\ref{neutrinoE}), showing that a field's contribution to dark energy is not a UV problem. The central question thus is uncovered: how heavy-field contributions are suppressed. This finding indicates that the current QFT understanding is insufficient, necessitating new mechanisms beyond conventional treatments. The microscopic result cannot be directly identified with the macroscopic cosmological constant, implying a scale separation that, in the Wilsonian framework \cite{Wilson:1971bg}, decouples microscopic zero-point energy from the macroscopic cosmological constant. We therefore adopt an effective description based on this scale decoupling, taking the total zero-point energy of free field modes as the natural baseline. The physical vacuum energy is then defined as the deviation from this baseline, caused by boundary conditions or interactions. We propose that dark energy originates from the cumulative effect of vacuum fluctuations mediated by carrier fields at the characteristic scale $\mu_\Lambda^{}$. Acting as a filtering scale, $\mu_\Lambda^{}$ selects which microscopic vacuum fluctuations can coherently build up to produce a macroscopic deviation from the baseline, such that fields with $m_i \lesssim \mu_\Lambda^{}$ contribute actively, while those with $m_i \gg \mu_\Lambda^{}$ decouple.

The above picture is captured by the ripple description (Fig. \ref{ripple1}): just as coherent surface ripples (not deeper water fluctuations) cause net water displacement, a net physical contribution requires vacuum fluctuations to achieve macroscopic coherence at the scale set by $\mu_\Lambda^{}$ (1/$\mu_\Lambda^{}$). This coherence is described by a fluctuation factor structuring the carrier field across scales, in analogy to the Boltzmann factor of energy quanta in thermal radiation. Given the Hubble parameter $H_0 \approx$ 70 km s$^{-1}$ Mpc$^{-1}$ and dark energy at 71\% of the critical density $3 H_0^2 / 8 \pi G$, the dark energy density is about $(2.3 ~ \mathrm{meV})^4$, corresponding to the scale $\mu_\Lambda^{}$ in the meV range by naturalness, indicating sensitivity to neutrino fields, especially to neutrino masses.\footnote{Other proposals for neutrino contributions include varying relic neutrino masses \cite{Fardon:2003eh,Fardon:2005wc,Takahashi:2005kw}, slow-roll quintessence with an extended neutrino sector \cite{Hung:2000yg,Barbieri:2005gj}, and cosmological axion-neutrino connection \cite{deVega:2023fvw}, etc.} Moreover, $\mu_\Lambda^{}$ should be slightly larger than the typical neutrino mass to yield a positive dark energy density. Considering Eq. (\ref{neutrinoE}) and dark energy density, a typical neutrino mass is $m_\nu^{\mathrm{tp}} \sim 2.3\times (32\pi^2)^{\frac{1}{4}} \approx$ 9.7 meV (here $|\log \frac{m_i^2}{\mu_\Lambda^2}|^{\frac{1}{4}}\sim 1$). As a rough prediction, there should be a neutrino with a mass $\sim$ 10 meV.

\begin{figure}[htbp!]
\includegraphics[width=0.39\textwidth]{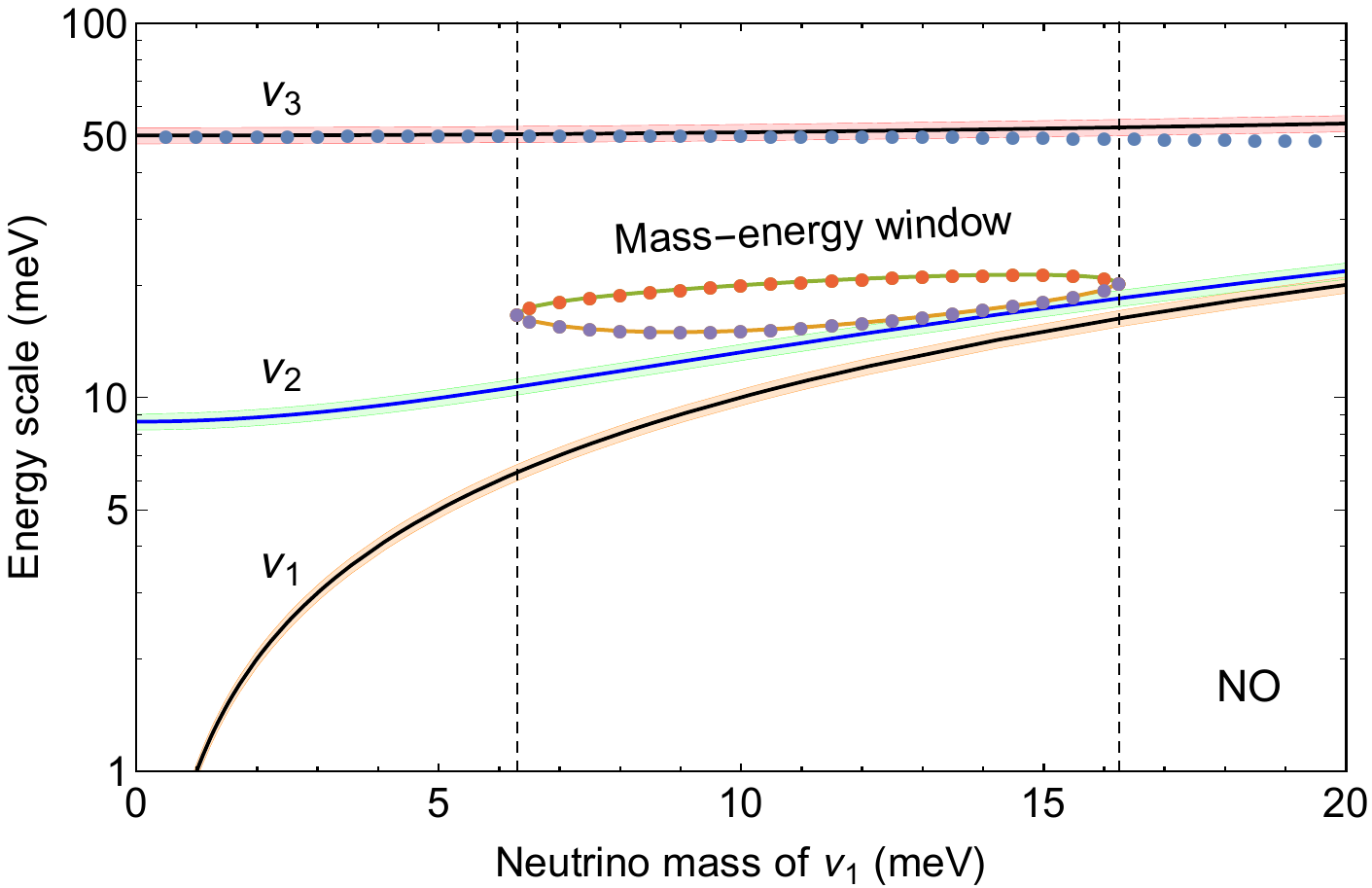}
\caption{The normal ordering neutrinos with the dark energy density $(2.3 ~ \mathrm{meV})^4$. The solid curves are masses of neutrinos with the mass of $\nu_1$ as the input, and the bands with a 5\% change in neutrino mass are plotted for a convenient naturalness estimation. The dotted curves are the values of $\mu_\Lambda^{}$ required by the dark energy density with neutrino masses inputted. The region between the two dashed lines is the neutrino mass window.}
\label{n-no}
\end{figure}

\begin{figure}[htbp!]
\includegraphics[width=0.39\textwidth]{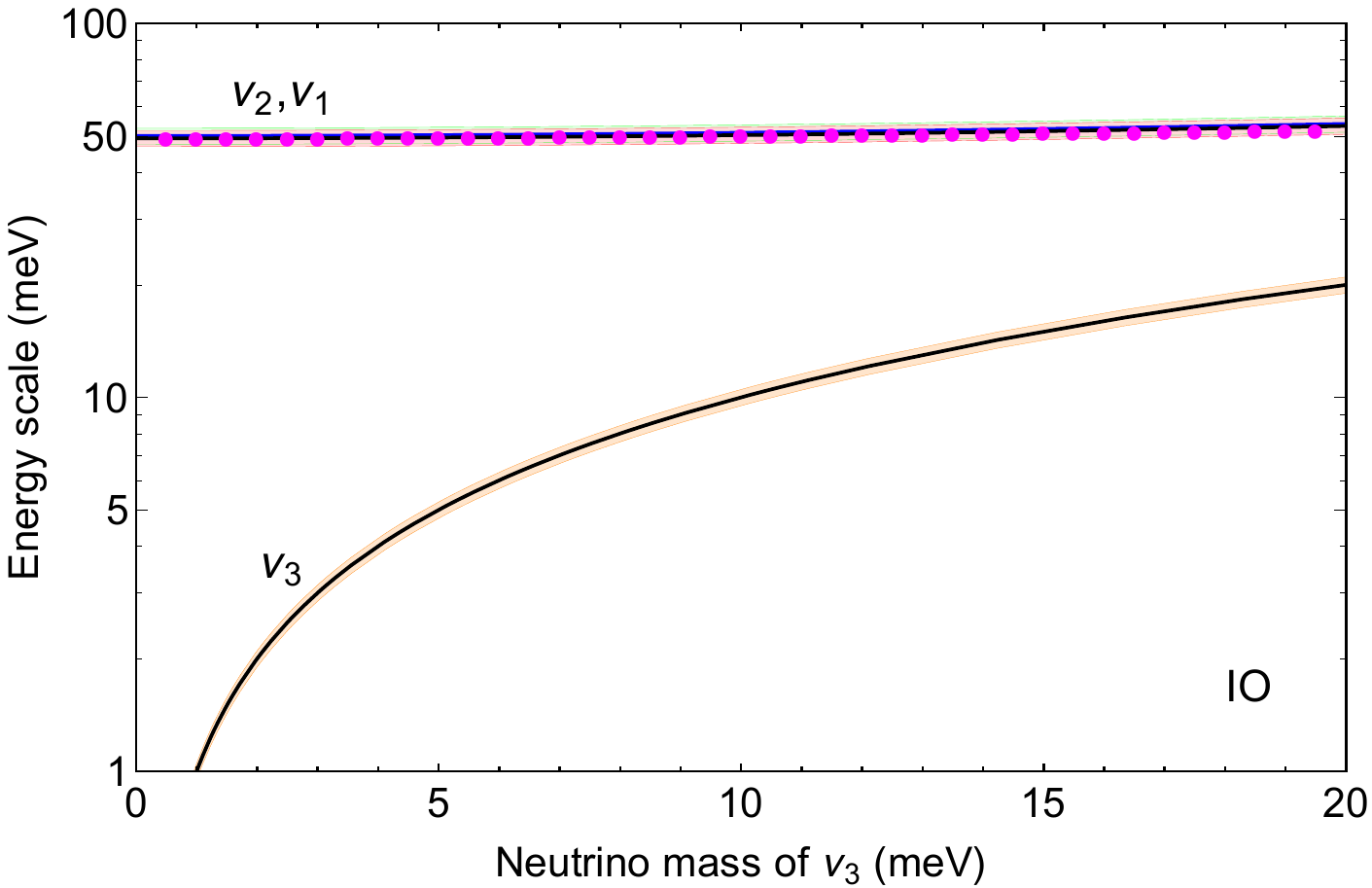}
\caption{The inverted ordering neutrinos with the dark energy density $(2.3 ~ \mathrm{meV})^4$. The solid curves are masses of neutrinos with the mass of $\nu_3$ as the input. The dotted curve is the values of $\mu_\Lambda^{}$ required by the dark energy density.}
\label{n-io}
\end{figure}

Now we turn to neutrino masses. The masses of three neutrinos are $m_1$, $m_2$ and $m_3$, and the mass differences between them can be extracted by neutrino oscillations. The mass-splitting results \cite{ParticleDataGroup:2024cfk} are $\Delta m_{21}^2 \approx 7.41 \times 10^{-5} \mathrm{eV}^2$, $\Delta m_{31}^2 \sim \Delta m_{32}^2 \approx 2.437 \times 10^{-3} \mathrm{eV}^2$ (normal ordering) and $\Delta m_{31}^2 \sim \Delta m_{32}^2 \approx -2.498 \times 10^{-3} \mathrm{eV}^2$ (inverted ordering). Considering the naturalness with an effective mass $m_\nu^{\mathrm{tp}}$, the quasidegenerate spectrum with $m_1 \simeq m_2 \simeq m_3 \gg \sqrt{|\Delta m_{32}^2|}$ is not favored. It is possible to give an estimate on the mass spectrum of neutrinos. Here we consider a specific implementation of the ripple description by a test fluctuation factor $e^{- m_i^2/\mu_\Lambda^{2}}$ in Gaussian distribution (see the \hyperref[appx]{Appendix}) to characterize the active contribution of a field, and $g_i$ together with this factor can be taken as an effective active degrees of freedom $g_i^\ast$ with $g_i^\ast = g_i e^{-m_i^2/\mu_\Lambda^{2}}$ ($\rho_\mathrm{vac} = \sum_i (-1)^{2 j_i} g_i^\ast \frac{m_i^4}{64 \pi^2} \log \frac{m_i^2}{\mu_\Lambda^2}$). For a field with a mass $m_i \gg \mu_\Lambda^{}$, its contribution to the dark energy density is negligible due to the tiny value of $g_i^\ast$. The hierarchical spectra of neutrinos for normal ordering and inverted ordering with the required dark energy density are shown in Fig. \ref{n-no} and Fig. \ref{n-io}, respectively. In Fig. \ref{n-io}, considering the naturalness of $\mu_\Lambda^{}$, the inverted ordering is not favored. For the normal ordering shown in Fig. \ref{n-no}, there is a region that neutrino masses at given $\mu_\Lambda^{}$ can naturally produce the physical vacuum energy required by the dark energy density, i.e., the parameter spaces forming a mass-energy window. Likewise, the neutrino mass window set by dark energy is 6.3 meV $\lesssim m_1 \lesssim$ 16.3 meV, 10.7 meV $\lesssim m_2 \lesssim$ 18.4 meV, 50.5 meV $\lesssim m_3 \lesssim$ 52.7 meV, and the total neutrino mass is 67.5 meV $\lesssim m_1+m_2+m_3 \lesssim$ 87.4 meV. This mass window can be tested by the future experiments.

Additionally, there is an inconsistency in the Hubble constant $H_0$ from different experiments. A slightly higher $H_0 = (73.04 \pm 1.04$) km s$^{-1}$ Mpc$^{-1}$ from the nearby universe \cite{Riess:2016jrr,Riess:2021jrx} and a slightly lower $H_0 = (67.4 \pm 0.5$) km s$^{-1}$ Mpc$^{-1}$ from the early universe \cite{Planck:2018vyg} correspond to dark energy densities of about $(2.37 ~ \mathrm{meV})^4$ and $(2.24 ~ \mathrm{meV})^4$, respectively. A slowly running $\mu_\Lambda$ (i.e., a running cosmological constant) could in principle affect the CMB-inferred $H_0$. The Hubble tension remains an open question \cite{Freedman:2024eph}, and a quantitative analysis is beyond the scope of this work.

\section{Conclusion and Discussion}
\label{sec:Con}

In this paper, the power-law divergence of vacuum energy has been addressed, and the UV-free scheme shows it is not a problem in the UV region. Freed from the veil of UV divergence, the core question becomes how unknown mechanisms suppress heavy-field contributions, a new open question. This work thereby recasts the cosmological constant problem from a fine-tuning puzzle into a defined physical question, thus clarifying the actual nature of the problem. In fundamental physics, such a redefinition is itself a necessary first step, while its full resolution is the next goal.

A new framework comes with its necessary shift in perspective.\footnote{This transformation responds to a paradigm crisis in QFT over the UV divergence problem: renormalization's ``subtracting infinity from infinity" induces the fine-tuning problem, undermining its coherence. The proposed remedies lack experimental support and deviate from naturalness, thereby forcing a shift to paradigms such as intrinsic finiteness. Such a shift manifests a conceptual incommensurability, marking a rupture in presuppositions rather than mere technique. Definitive progress is therefore defined not by the criteria within the traditional framework, but by constructing a new, logically consistent system that dissolves its predecessor's core impasses and yields testable predictions.} Accordingly, we address two recurrent questions (also noted in Ref. \cite{Cline:2024klo}): why $\mu_\Lambda^{}$ lies at the meV scale, and why contributions from heavy fields are neglected. To the first question, there is currently no derivation of $\mu_\Lambda^{}$ from fundamental principles, similar to values of SM particle masses and couplings. Considering the naturalness of an energy scale that characterizes the macroscopic effect of vacuum energy, $\mu_\Lambda$ should not be too high; equivalently, its associated length scale should not be too small. Here $\mu_{\Lambda}^{}$ is an input parameter, and its parameter space is derived from the observed dark energy density guided by naturalness. Theoretically, this scale may link to the inflationary field or the origin of neutrino masses. A connection to neutrinos is of particular interest, as their mass scale would bridge quantum vacuum physics and cosmology. To the second question, this work explores the $\mu_\Lambda$-characterized scale decoupling of vacuum energy from microscopic quantity to macroscopic effect, using statistical physics to yield the decoupling of heavy fields. Consequently, the dark energy problem could be minimally addressed by introducing the new scale $\mu_\Lambda^{}$ into the SM.

The scale $1/\mu_\Lambda$ provides the quantum basis for the universe's homogeneous, isotropic accelerated expansion. Acting as a universal filter, it selects vacuum fluctuations at this scale to coherently form dark energy, yielding uniform density. Homogeneity follows from this universality. All such fluctuations are uniformly selected and superposed. Isotropy stems from the quantum vacuum's lack of preferred direction, resulting in a spherical macroscopic average. Thus, the cosmological principle's classical features emerge from macroscopic coherence at $1/\mu_\Lambda$. Matter's gravitational effects then superimpose on this background, with vacuum energy and matter co-evolving as a fundamental cosmic component. In addition, the dark energy density is so small that its curvature effect (set by the Hubble scale) on local QFT is negligible in vacuum energy calculations.

To account for the dark energy density, a neutrino with a typical mass around 10 meV is expected in the ripple description. Considering the naturalness at the scale $\mu_\Lambda^{}$, the normal hierarchical spectrum of neutrinos is preferred, and there is a mass-energy window for the neutrino mass spectrum and the dark energy density. The neutrino mass window set by dark energy can be examined by the future experiments. Moreover, for UV-divergence inputs in QFT, there may be another route besides the routes mentioned in this article. All roads lead to the same goal. With the joint explorations of multiple routes, we hope it will be possible to get close to the harmonious and unified physical structure hidden beneath UV divergences. We look forward to more explorations.

\section*{Acknowledgements}

The author thanks James M. Cline for useful discussions. This work was partly supported by the fund of SWUST under the contract no. 23cxtd59.

\appendix*

\section{Fluctuation factor}   \label{appx}

For a mass $m$ carrier field, the $\mu_\Lambda^{}$-characterized fluctuation factor describes macroscopic effects of cumulative vacuum fluctuations from microscopic scales. Its potential form is derived as follows. The first is basic assumptions, i.e., \\
(a) Independence assumption. Consider a system with $N$ independent random fluctuations $\{q_i\}_{i=1}^N$, satisfying: Zero expectation, with $E [q_i] = 0$; Finite variance, with $Var(q_i) = \sigma_i^2$ (bounded fluctuation strength); Statistical independence between random variables $q_i$.  \\
(b) Linear superposition. The total cumulative effect $X$ is a linear superposition of individual fluctuations with $X = \sum_{i=1}^N a_i q_i$, where $a_i$ being weight coefficients, and the system exhibits no significant nonlinear coupling.

Next is the derivation via the characteristic function. For each $q_i$, its characteristic function $\phi_i(\tau)$ = $E[e^{i \tau q_i}]$ expands to $\phi_i(\tau)$ = $1 + i \tau E [q_i] - \frac{\tau^2}{2} E [q_i^2] + \cdots \approx 1 - \frac{\sigma_i^2 \tau^2}{2}$, where $\cdots$ denotes higher-order terms. By independence, the characteristic function of $X$ is the product of individual characteristic functions, $\phi_X(\tau) = \prod_{i=1}^N \phi_i(a_i \tau) \approx \prod_{i=1}^N (1 - \frac{a_i^2 \sigma_i^2 \tau^2}{2} )$. Taking the logarithm and linearizing (using $\ln(1+x) \approx x$ for $x \to 0$), the result is $\ln \phi_X(\tau) \approx -\frac{\tau^2}{2} \sum_{i=1}^N a_i^2 \sigma_i^2 $ $\Rightarrow$ $\phi_X(\tau) = \exp (-\frac{\sigma_X^2 \tau^2}{2} )$, where $\sigma_X^2 = \sum_{i=1}^N a_i^2 \sigma_i^2$. This is the characteristic function of a Gaussian distribution. In this paper, we assume that the cumulative effect $X$ of vacuum fluctuations manifests macroscopic effects through carrier fields. Specifically, the fluctuation factor $f(m)$ is set by $\phi_X(\tau)$, and it can be obtained via inverse Fourier transform of $\phi_X(\tau)$ with a matching condition $X \to m$, that is, $f(m) = A \frac{1}{2\pi} \int_{-\infty}^\infty e^{-i\tau m} \phi_X(\tau) \, d\tau = \frac{A}{\sqrt{2\pi \sigma_X^2}} \exp (-\frac{m^2}{2\sigma_X^2} )= B \exp (-\frac{m^2 \pi B^2}{A^2} )$, with $A$ being a coefficient, and $B=\frac{A}{\sqrt{2\pi \sigma_X^2}}$. Considering photon field as the carrier field for normalization, one has $f(0) =1$, and thus $B=1$ is obtained. Taking $A^2 / \pi = 2\sigma_X^2 = \frac{1}{k} \mu_\Lambda^2$, the fluctuation factor is written as $f(m) = e^{- k m^2/\mu_\Lambda^{2}}$, and $k \sim 1$ can be adopted for $\mu_\Lambda^{}$-characterized fluctuations.

From another perspective, the macroscopic effect of physical vacuum energy requires a long-range description with characteristic scale $\sim 1/\mu_\Lambda$. This reflects a general concept of statistical physics, namely that microscopic energies are filtered into macroscopic effective quantities. A classic example is gas pressure, which arises from the statistically averaged translational kinetic energy of molecules, but not from their rest energy. Hence, macroscopic observables are scale-selected subsets of microscopic energies.

Applying this concept to dark energy, we interpret the dark energy density as a macroscopic observable governed by a $\mu_\Lambda$-scale filter. The microscopic zero-point energy of a quantum field is enormous, yet only fluctuation modes coherent over $\sim 1/\mu_\Lambda$ contribute effectively ($X \to 1/\mu_\Lambda$, with $A^2 / \pi = 2\sigma_X^2 = \frac{1}{k m^2}$). For a field with $m \gg \mu_\Lambda$, its coherence length (the Compton wavelength $\lambda_c \sim 1/m$) is far smaller than $1/\mu_\Lambda$. High-frequency fluctuations of a heavy field cannot maintain phase coherence over this macroscopic scale. Consequently, a macroscopic volume of size $\sim 1/\mu_\Lambda$ contains a large number of independent fluctuation patches with random phases. Their superposition leads to cancellation under macroscopic averaging, resulting in exponential suppression. Therefore, the cumulative effect of a heavy field is strongly suppressed. The observed small dark energy density does not contradict huge microscopic energies; rather, it shows that the cosmos is sensitive only to modes passing the decoupling filter $\lambda_c \gtrsim 1/\mu_\Lambda$. This phase-statistical filtering provides a statistical explanation for the smallness of dark energy density.


\end{document}